\documentclass[twocolumn,showpacs,preprintnumbers,amsmath,amssymb,groupaddress,superscriptaddress]{revtex4}

\usepackage{graphicx}
\usepackage{dcolumn}
\usepackage{bm}

\begin{document}

\preprint{APS/123-QED}

\title{Efficient Response to Cascading Disaster Spreading}

\author{Lubos Buzna}
\email{buzna@vwi.tu-dresden.de}
\affiliation{Dresden University of Technology, Andreas Schubert Str. 23, 01062 Dresden, Germany}
\affiliation{University of Zilina, Univerzitna 8215/5, 01026 Zilina, Slovakia}

\author{Karsten Peters}%
\affiliation{Dresden University of Technology, Potthoff-Bau, Hettner Str. 1-3, 01062 Dresden, Germany}

\author{Hendrik Ammoser}%
\affiliation{Dresden University of Technology, Andreas Schubert Str. 23, 01062 Dresden, Germany}

\author{Christian K\"uhnert}%
\affiliation{Dresden University of Technology, Andreas Schubert Str. 23, 01062 Dresden, Germany}

\author{Dirk Helbing}%
\affiliation{Dresden University of Technology, Andreas Schubert Str. 23, 01062 Dresden, Germany}
\affiliation{Collegium Budapest-Institute for Advanced Study, Szenth\'aroms\'ag u. 2, 1014 Budapest, Hungary}

\date{\today}

\begin{abstract}
We study the effectiveness of recovery strategies for a dynamic model of failure spreading in networks. These strategies control the distribution of resources based on information about the current network state and network topology. In order to assess their success, we have performed a series of simulation experiments. The considered parameters of these experiments are the network topology, the response time delay and the overall disposition of resources. Our investigations are focused on the comparison of strategies for different scenarios and the determination of the most appropriate strategy. The importance of prompt response and the minimum sufficient quantity of resources are discussed as well.
\end{abstract}


\maketitle

\section{INTRODUCTION}

The efficient distribution of resources is a challenging problem relevant for many types of complex networks. The examples include social networks \cite{Davidsen}, power transmission grids \cite{Motter}, communication systems \cite{Rosato, Newman}, and road infrastructures \cite{Kalapala}. Physicists, in the past years, have studied their structure and considerably contributed to the understanding of processes going on these networks. The efficient immunization against the epidemic spreading of diseases and strategies for failure prevention are important topics with many practical implications in real systems. Scientists have recently demonstrated the benefits of targeted \cite{Dezso, Satorras, Goldenberg} and acquaintance immunization \cite{Cohen} in scale-free networks, have studied the applicability of "flooding" dissemination strategies based only on local information \cite{Stauffer2}, and have proposed efficient strategies for eliminating cascading effects in networks \cite{Motter2, Schafer}.

In contrast to these works we would like focus on interdependent systems and on the spreading dynamics of disastrous events between the networked components. Disastrous events are bothering mankind from the earliest days. The ability to recover the functionality of damaged infrastructures promptly is crucial for survival and determines, whether the affected areas will overcome the consequences of catastrophe or not. Emergency response and recovery  call for external resources, which are limited and, therefore, have to be deployed as efficiently as possible. The question how to effectively distribute resources in order to fight disasters best has already been addressed by many researchers. As examples, we mention the redistribution of medical material \cite{Tuson},  the mitigation of large-scale forest fires \cite{Fiorucci}, and the fighting of floods \cite{Altmann}.

An experimental study of disasters under real world conditions is almost impossible, and therefore, mathematical and computer models are often very helpful tools to extend human knowledge. However, the complexity of systems struck by disasters does not allow one to model the interactions of all involved entities and processes in detail. Therefore, we have to capture them by an appropriate generic model.  Disastrous events are often characterized by cascading failures \cite{Helbing2} propagating in the system due to the causal dependencies between system components. These casual dependencies result from structural and functional interdependencies and can be modeled by directed networks. Note that there were several attempts to quantify such networks for particular cases, using interaction network approaches \cite{Helbing} or fuzzy cognitive maps \cite{Papageorgiou}. Loops in these networks are crucial, since the amplification of negative effects through the loops may considerably deteriorate the situation. Such loops are sometimes called "vicious circles".

The above mentioned view of disasters has led us to the formulation of a general spreading model of failures in networks \cite{Buzna}. To assess the importance of the availability of information about the network on the efficiency of disaster recovery, in this paper we will study the effect of different protection strategies. These strategies are based on different information evaluation and control the distribution of resources over the system components. As parameters in our model, we consider the overall quantity of resources $R$, the recovery time delay $\tau_{start}$, and the network topology $M_{ij}$. As our simulations did not give qualitatively different results for varying link weights $M_{ij}$, we will not discuss the case of heterogeneous network links in this paper. Our presented simulation results rather focus on the average efficiency of the considered strategies and on the "worst-case" scenario, which is given by the most "unfriendly" realization of all random parameters. 

Our paper is organized as follows: Sec. \ref{sec:mod} presents our mathematical model of disaster spreading. In Sec. \ref{sec:mob}, we describe the mobilization process of resources. Disaster recovery modeling issues and protection strategies are discussed in Sec. \ref{sec:cri}, while the results of our computer simulations are presented in Sec. \ref{sec:res}. To conclude this paper, Sec. \ref{sec:con} summarizes the most important findings and outlines possible directions of future research.

\section{MODELING THE DYNAMICS OF DISASTER SPREADING}
\label{sec:mod}

In this section, we briefly summarize our model of disaster spreading originally proposed in \cite{Buzna}. The model is based on a graph ${\bf \mathcal{G}} =  (\mathcal{N},\mathcal{S})$ of interconnected system components $ i \in \mathcal{N} = \{ 1, \dots,n \}$. The directed links $ (i,j) \in \mathcal{S}$, with $i,j \in \mathcal{N}$, represent structural and functional dependencies between the components. The state of a node $i$ at time $t$ is described by a continuous variable $x_{i}(t) \in \mathbb R$, where $x_{i} = 0$ corresponds to a normal functioning of the component. The deviation from this state, caused by disturbances, represents the level of challenge of system component $i$. At the present stage of abstraction, we do not consider diverse functionalities of the components and we assume an additive impact of external disturbances coming from neighboring components.

Each real system exhibits a natural level of resistance to challenges. We reflect this tolerance by a special threshold $\theta_{i} > 0$ and assume that a node tends to fail, when the sum of all disturbances acting on it exceeds this value \cite{Altmann}. Rather than by a discontinuous step function we describe this by the sigmoidal function

\begin{equation}
\Theta(y) = \frac{1-\exp\left(-\alpha y \right)}{1 + \exp \left[- \alpha(y - \theta_{i})\right]},
\end{equation}
where $\alpha$ is a "gain parameter". 

The interactions between the components are quantified by the connection strengths $M_{ij}$ and by the link transmission time delays $t_{ij} > 0$. The overall dynamics of a node is then given by:

\begin{equation}
\frac{d x_{i}}{d_{t}} = -\frac{x_{i}}{\tau_{i}} + \Theta\left( \sum_{i \neq j} \frac{M_{ji} x_{j}(t - t_{ji})}{f(O_{j})}e^{-\beta t_{ji}}\right), 
\label{eq:node}
\end{equation}
where the first term on the right-hand side models the ability of component $i$ to recover from perturbations and the second term describes the superposition of all pertubative influences by adjacent nodes $j$ on node $i$. If ${x_{i} \neq 0}$, the recovery term tends to drive $x_i$ back to zero. The recovery rate $1/\tau_{i}$ characterizes the speed of the recovery process at node $i \in \mathcal{N}$. The function ${f(O_{j})=(aO_j)/(1 + bO_j)}$ introduces an additional weight to reflect that the impact of highly connected neighboring nodes is smaller, because their influence is distributed among many nodes and in this way "dissipated" \cite{footnote}. $O_j$ is the out-degree of node $j$ while $a = 4$ and $b = 3$ are fit parameters. 
\par
The disturbances, as they are transmitted over the links, can be strengthened or weakened by different factors like for instance the time delays or physical properties of the surrounding. The intensity of this process can, in our model, be controlled by the parameter $\beta$. In the experiments we have used $\beta = 0.025$, which corresponds to relatively weak damping of disturbances on links.

\subsection{Setup of Our Simulation Experiments}
\label{subsec:set}

Our simulation studies were performed for four types of directed networks representing different systems. Specifically, we have studied networks such as regular (grid) networks, random networks, scale-free networks, and small-world networks. 

Only regular (grid) networks were specified with bidirectional links. The directed scale-free networks were generated using the algorithm by Bollob\'as, Borgs, Chayes, and Riordan \cite{Bollobas}, where the attachment of new node is controlled by probabilities $\alpha_{1}$, $\beta_{1}$, $\gamma_{1}$ with ${\alpha_{1} + \beta_{1} + \gamma_{1} = 1}$ and by non-negative parameters  $\delta_{in}$ and $\delta_{out}$. These parameters have been set to ${\alpha_{1} = 0.1}$, ${\beta_{1} = 0.8}$, ${\gamma_{1} = 0.1}$, ${\delta_{in} = 2}$ and ${\delta_{out} = 2}$.

Small-world networks have been generated using the procedure described in Ref. \cite{Murai}. This procedure slightly generalized the generation of {\it undirected} small-world graphs proposed by Watts and Strogatz \cite{Watts}: In contrast to their original algorithm, we have randomly assigned directions to links, with probabilities for clockwise and counter-clockwise direction of 0.3 each, while a bidirectional link has been assumed with probability 0.4. Finally, a random rewiring procedure with rewiring probability $p = 0.3$ has been applied.

In addition, we have generated random networks of the Erd\"os-R\'enyi type. All networks have been generated in a way that the resulting average node degree was approximately 3.6. The grid network was organized in 25 rows each containing 20 nodes.

Throughout this paper, all computer-generated networks are composed of 500 nodes. Moreover, our homogeneous parameter settings assume that all $\theta_{i}=0.5$ and all $M_{ij} = 0.5$, where a link from node $i$ to $j$ exists, otherwise $M_{ij}=0$. The time delays $t_{ij}$ are $\chi^{2}$-distributed, where we have chosen ${\mu = 4}$ for the number of degrees freedom of the $\chi^{2}$-function. However, the distribution was stretched multiplying by factor 0.05 and shifted by adding the value 1.2 in order to get an average delay of ${\langle t_{ij} \rangle = 1.4}$.

\section{ MOBILIZATION OF RESOURCES}
\label{sec:mob}
\begin{figure}
    \begin{center}
    \includegraphics[angle=0,width=0.38\textwidth]{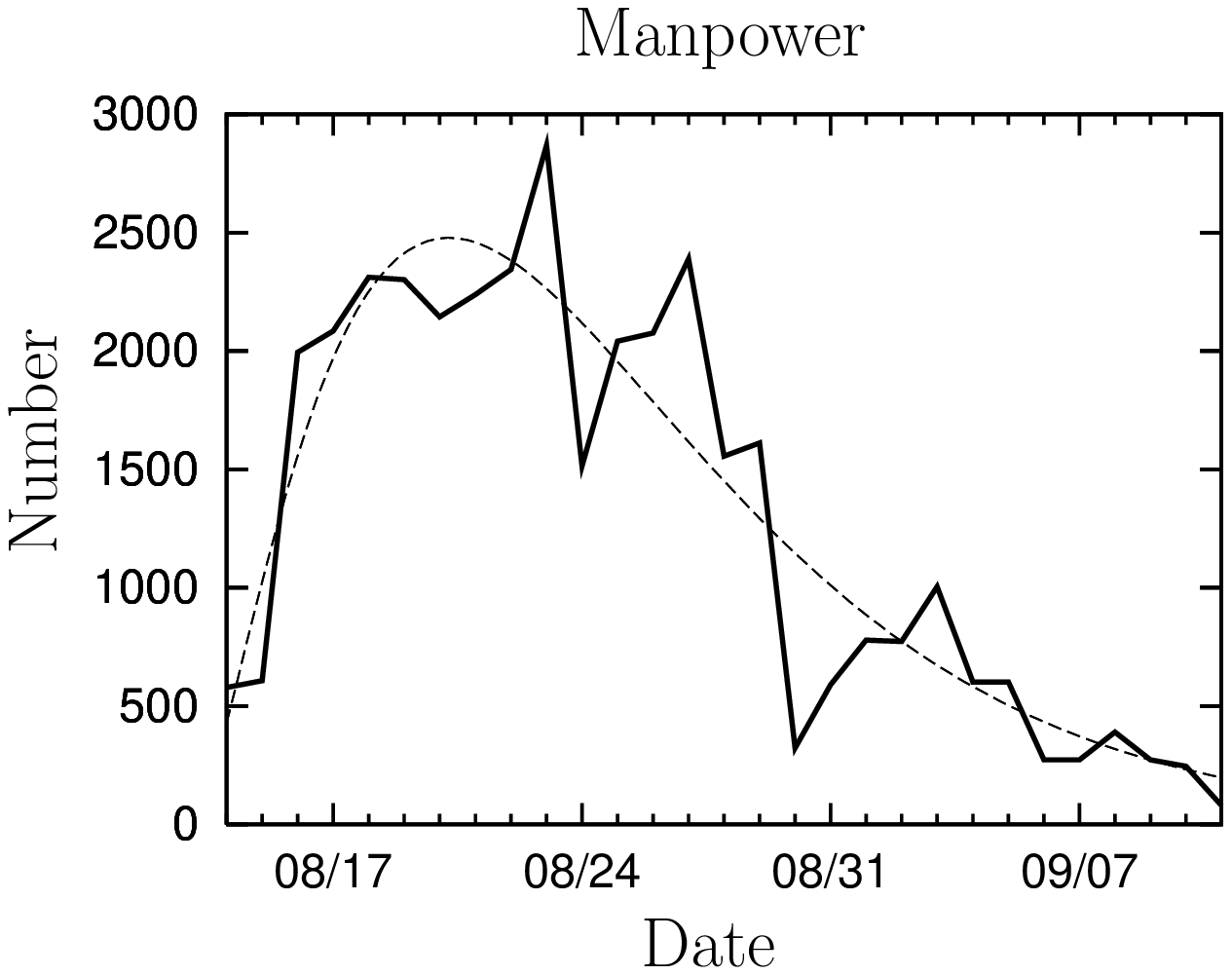}
    \includegraphics[angle=0,width=0.38\textwidth]{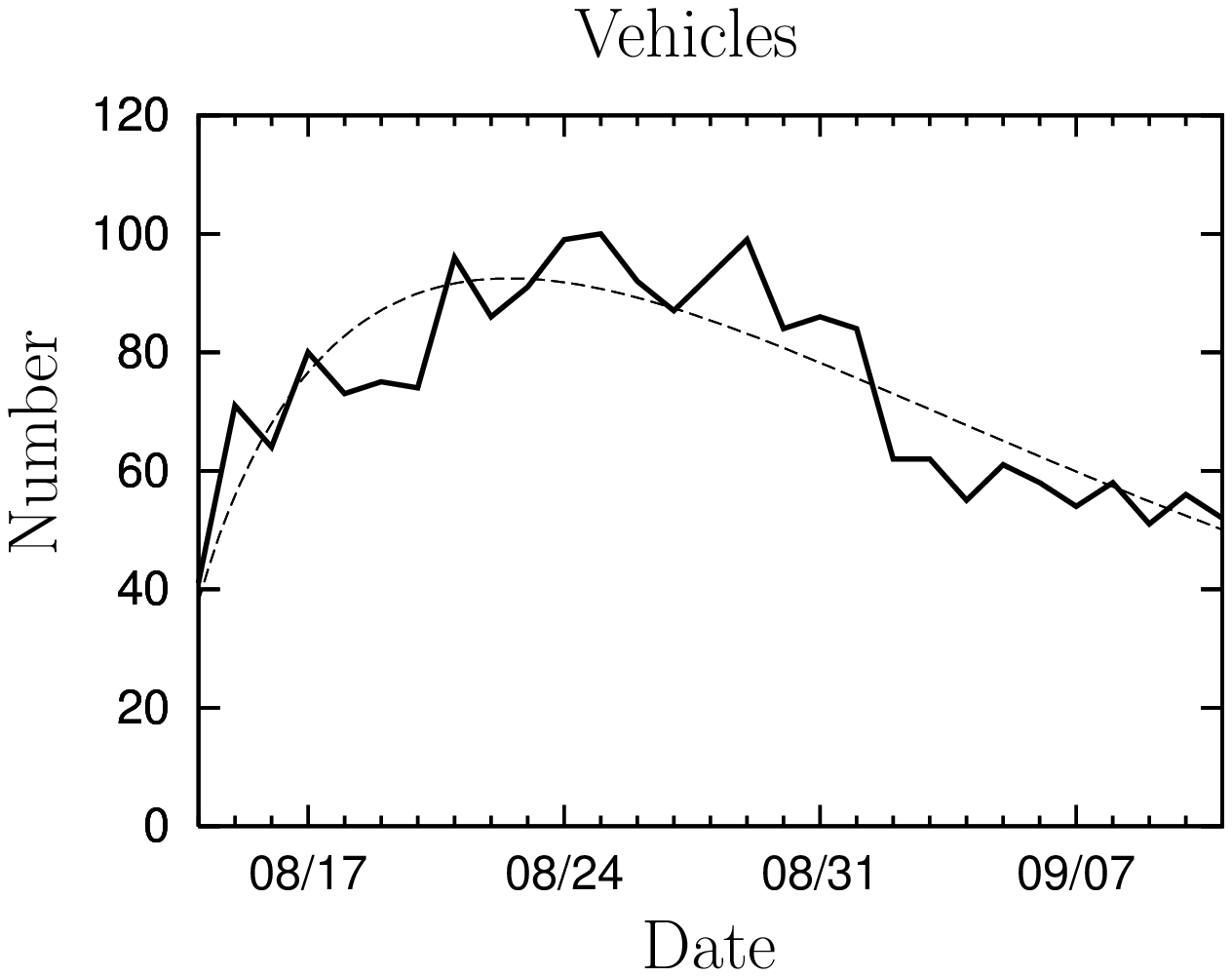}
  \end{center}
    \caption{ Manpower and number of  vehicles fighting the Elbe river flooding in August 2002 \cite{Kirchbach}. The dashed lines represent the approximation by the function $r(t) = a_{1} t^{b_{1}}e^{-c_{1}t}$. The best fit parameters for manpower (top) are $a_{1} =530$, $b_{1} = 1.6$, $c_{1} = 0.22$, while for vehicles (bottom) they are $a_{1} =41$, $ b_{1} = 0.66$, $ c_{1} = 0.069$.}
    \label{fig:data}
\end{figure}

Let us assume that the emergency forces and all material flows are entering the affected area continuously in time. This process can be modeled by a continuous function, which defines, how much resources have reached an affected area at time $t$. The shape of this function is an essential point of our model, because the prompt mobilization of resources a has strong influence on the efficiency of countermeasures \cite{Eubank}. Despite the frequent occurrence of disasters, we found only a few publications that provide a detailed information about the progress of mobilization in time. For example, Fig. \ref{fig:data} shows the manpower and vehicles, which were involved in the recovery activities to fight the Elbe river flooding in Germany in August 2002 \cite{Kirchbach}. Both curves are quantitatively similar and can be well approximated using the function $r(t) = a_{1} t^{b_{1}}e^{-c_{1}t}$, where $a_1$, $b_1$ and $c_1$ are fit parameters. The mobilization itself is represented by the growing part of the curve. To reflect the progress of mobilization of external resources in our simulations, we have used the approximate fit curve for manpower. Besides time progress of the mobilization, further important parameters are the overall quantity of external resources $R$ and the response time $t_{D}$. The response time is the time interval between the occurrence of the initial disastrous event and the first provision of resources. 
\par
The resources used for the recovery are assumed to be distributed in time according to the manpower data presented in Fig. 1. We have normalized the magnitude of this curve according the total amount of resources $R$, keeping its shape. The time period during which the distribution of resources takes place was set to half of the simulation time horizon, i.e. $50$ time steps.

\section{CRISIS MANAGEMENT AND DISASTER RECOVERY}
\label{sec:cri}

Disasters come mostly unexpected, and the first moments after their occurrence are characterized by a high uncertainty in the estimation of the overall impact. Crisis management coordinates the work of all emergency units and often has to take decisions based on scarce information. This requires a reliable organization in term of information flows \cite{Helbing3, Stauffer}, their evaluation and the choice of appropriate respose strategies.

To uncover what information is most important for efficient disaster response, we study here the properties of several recovery strategies, allocating the resources to  components based on different information. As first kind of information, let us consider the knowledge of the component's connectivity, i.e. the out-degrees and in-degrees of the nodes. This information allows one to uncover those components, which influence most other components and those which are easily vulnerable, because they have many in-going links. As second kind of information we assume that the locations and seriousness of malfunctions in the network are well-known. This information reflects the current level of node damage and allows one to prioritize the nodes which are more seriously damaged. Considering these two kinds of information, we have formulated the following recovery strategies $S_i$:

\begin{itemize}
\item[$S_1$] {\it uniform dissemination}, i.e. each node gets the same amount of resources,
\item[$S_2$] {\it out-degree based dissemination}, i.e. the resources are distributed over nodes proportionally to their out-degrees,
\item[$S_3$] {\it uniform reinforcement of challenged nodes}, i.e. all nodes ${i \in {\mathcal{N}}}$ with ${x_{i} > 0}$ are equally provided with resources,
\item[$S_4$] {\it simple targeted reinforcement of destroyed nodes}, i.e. damaged nodes (${x_{i} > \theta_{i}}$) are equally provided with resources with priority, while challenged nodes ($x_{i} > 0$) are uniformly reinforced if no damaged nodes exist,
\item[$S_5$] {\it simple targeted reinforcement of highly connected nodes}, i.e. a fraction $q$ of highly connected nodes is uniformly provided with resources by using the fraction $k$ of all resources, while the remaining resources are applied according to strategy $S_4$,
\item[$S_6$] {\it out-degree based targeted reinforcement of destroyed nodes}, i.e. application of strategy $S_4$, but with a distribution of resources proportional to the out-degrees of nodes rather than a uniform distribution.
\end{itemize}

Equation (\ref{eq:node}) represents the mitigation activities in the nodes by the recovery rates $1/{\tau_{i}}$. 
It models a situation without additional external forces sent to challenged system components to perform mitigation actions.  Thus, at the beginning it is assumed that the mitigation activities are weak ($1/\tau_{i} =  1/\tau_{start} = 0.25$), because they are based only on internal resources. If these internal resources are not sufficient to cope with the evolving disaster, external resources have to be mobilized. The assignment of external resources to a node is assumed to increase the recovery rate $1/\tau_i(t)$ of a node according to

\begin{equation}
\frac{1}{\tau_{i}(t)} = \frac{1}{(\tau_{start} - \beta_{2}) e^{-\alpha_{2} R_{i}(t)} + \beta_{2}}.
\label{eq:healing}
\end{equation}
Our model assumes that, once resources have been assigned to a node, they will remain at the selected node and are not reassigned again. In Eq. (\ref{eq:healing}), the cumulative amount of resources assigned to node $i$ is denoted by $R_i$. The formula reflects the fact that each new unit of resources has a smaller effect than the previous one, which is due to the decreasing efficiency of recovery activities, when the concentration of forces grows. These effects are well-known and may be explained by increasing efforts for communication and the coordination of forces. The influence of this effect is represented by the fit parameter $\alpha_2 = 0.58$. The parameter $1/\beta_2 = 5$ defines an upper bound of the recovery rate.
\par
When developing formula (\ref{eq:healing}), we have required the following: 
\begin{enumerate}
\item
Resources have only positive influence  on the state of the node. In other words, the function $1/\tau_{i}$ should grow monotonously with the parameter $R_i$.
\item
When there are no resources applied in node $i$ i.e., $R_i  = 0$, then $1/\tau_i  = 1/\tau_{start} > 0$.
\item
Finally, we expect a limited speed of recovery process. In fact, for $R_i \rightarrow \infty $ we have $1/\tau_{i}(t) = 1/\beta_{2}  > 0$.
\end{enumerate}
Formula (\ref{eq:healing}) obeys all three conditions, and we expect qualitatively similar results for all continuous functions satisfying these conditions.

\section{RESULTS OF OUR SIMULATION EXPERIMENTS}
\label{sec:res}
\begin{figure}
    \begin{center}
    \includegraphics[angle=0,width=0.38\textwidth]{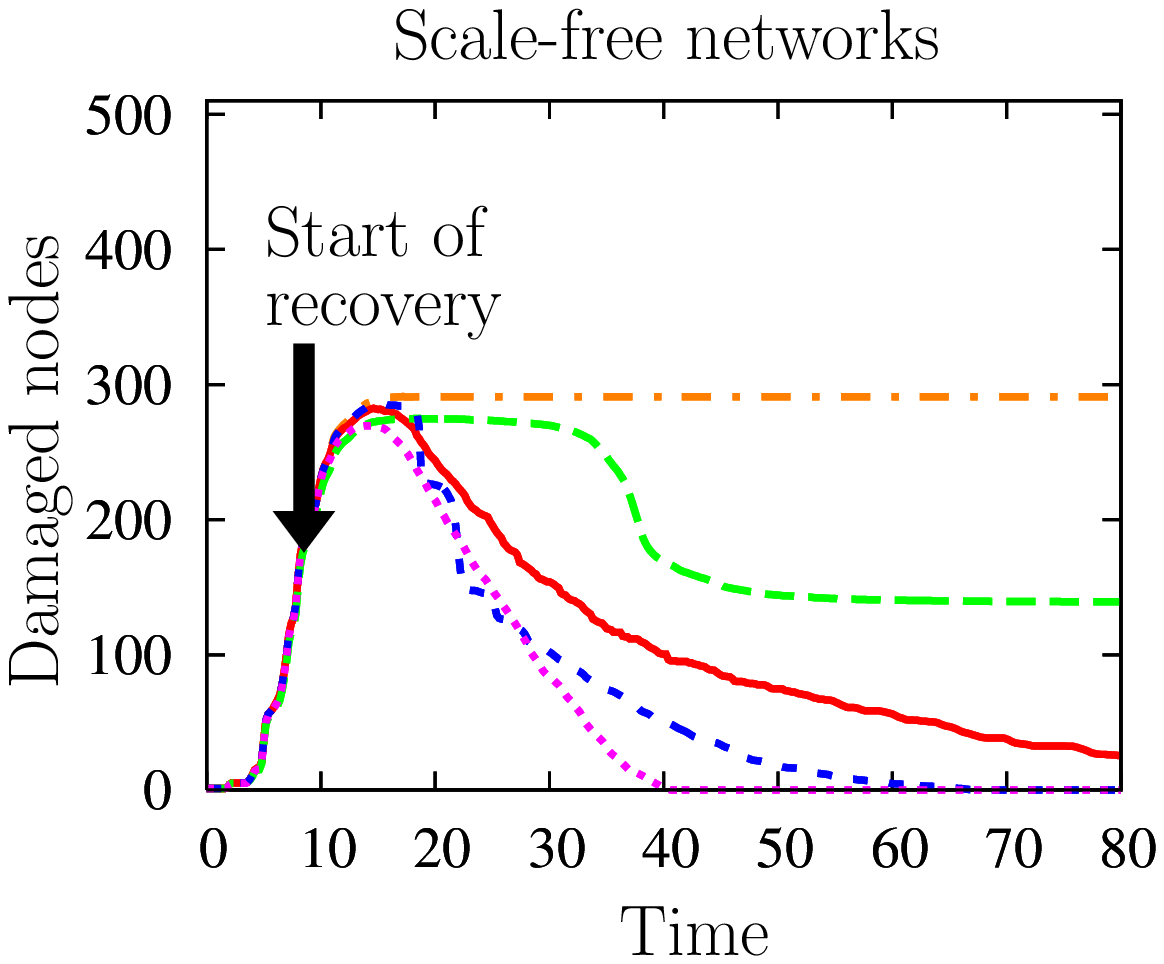}
    \includegraphics[angle=0,width=0.38\textwidth]{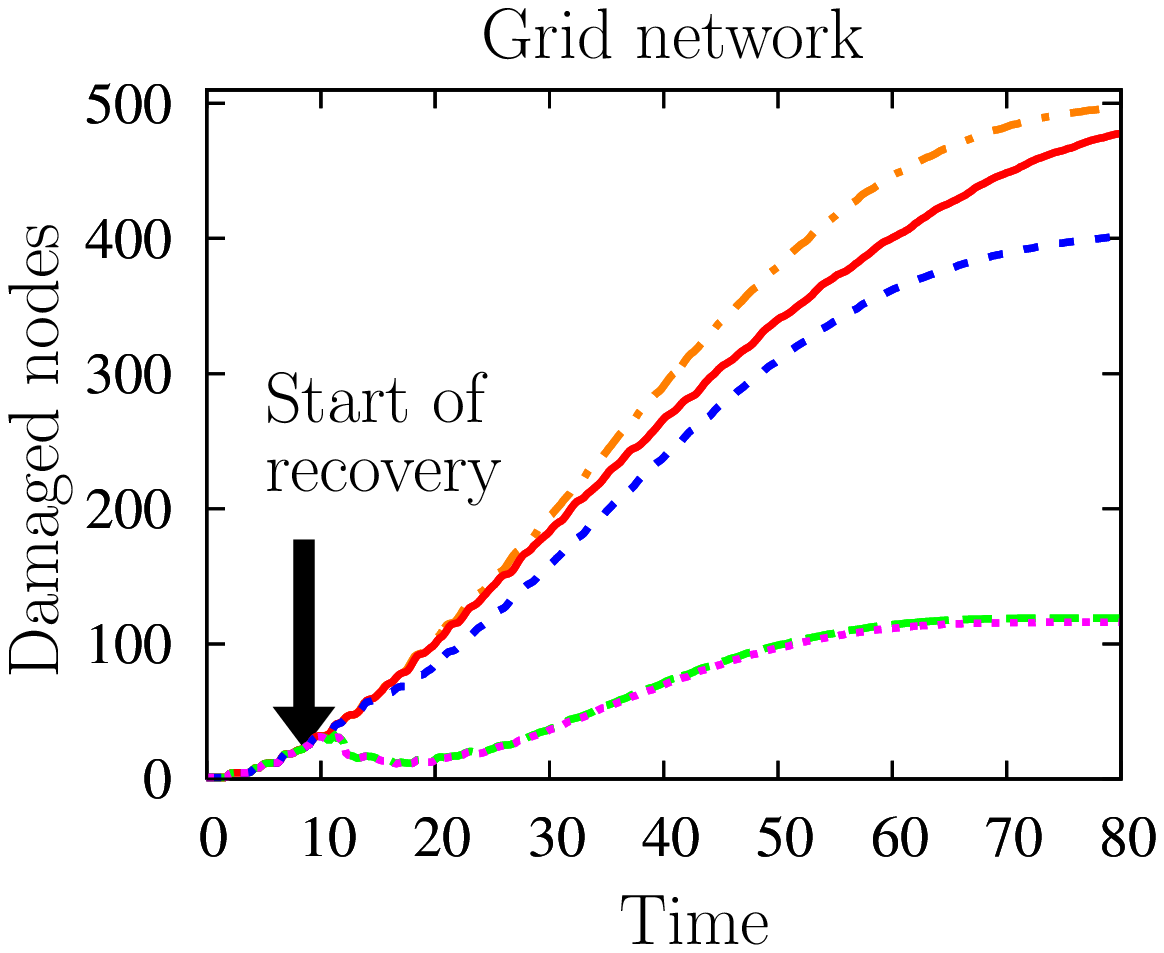}
  \end{center}
  \caption{(Color online) Average number of damaged nodes ($x_i > \theta_i$) for scale-free networks and a regular grid networks, applying different protection strategies. Dashed-dotted line: no disposition of resources for recovery. Solid line: strategy $S_{3}$. Long-dashed line: strategy $S_{4}$. Short-dashed line: strategy $S_{5}$. Dotted line: strategy $S_{6}$. The value of the response time was set to $t_{D} =  8$ and the overall disposition of resources to $R =1000$ (apart from the dashed-dotted line, where R = 0).}
\label{fig:damage}
\end{figure}

We have extensively studied the properties of protection strategies by means of computer simulations. Due to the existence of random parameters, such as $t_{ij}$, the results of the simulation experiments varied with the realizations of the random variables. Experiments started at time $t = 0$, when the $x_i$ variable of one randomly selected node $i$ was set to the value $\tau_{start}$ for 10 time units. Figure \ref{fig:damage} shows an example how the average number of damaged nodes than develops in the course of time. The existence of hubs causes that the perturbation propagates much faster in scale-free networks than in grids, but on the other hand, the protection strategies work more efficiently, when they can focus on these highly connected nodes. To assess the behavior of our model, we have evaluated the most unpleasant scenario, which occurs when we consider the most unfriendly realization of the random parameters. One possible characteristics, which reflects this "worst-case" scenario, is the dependence of the minimal quantity of resources $R_{min}$ required to recover the network on the response time delay $t_{D}$. It defines a success threshold for each considered strategy. Except for this, we have evaluated the average damage of the respective network. Therefore, all experiments have been performed with the same simulation time horizon ($t_{S} = 100$).

\subsection{Worst-Case Scenario}
\label{subsec:wor}

In this subsection we determine the minimum required resources $R_{min}$ as a function of response strategy and the network topology, and we study how $R_{min}$ changes when the response time delay increases. $R_{min}$ is the minimum quantity of resources which guarantees the complete recovery of the network for each particular scenario. We estimate this quantity by performing a huge amount of numerical calculations separately for each studied network. In each simulation run, the location of the initial disturbance and the time delays $t_{ij}$ are randomly varied. To obtain $R_{min}$, we use the bisection method.

\begin{table}
\caption{\label{tab:tab1}Values of $R_{min}$ obtained for strategies $S_{1}$ and $S_{2}$. The rows correspond to the different network types: square grid (GR), small-world networks (SW), Erd\"os-R\'enyi networks (ER), and scale-free networks (SF). The variance in data was obtained by $t_D$ moving over values $0, ..., 15$.}
\begin{ruledtabular}
\begin{tabular}{lcccccr}
&\multicolumn{2}{c}{$S_{1}$}&\multicolumn{2}{c}{$S_{2}$}\\
\hline
GR & 1954 &$\pm$ 25 & 2223 &$\pm$ 66\\
SW & 1861 &$\pm$ 8  & 1993 &$\pm$ 4\\
ER & 1701 &$\pm$ 4  & 1521 &$\pm$ 7\\
SF & 2203 &$\pm$ 0  & 1205 &$\pm$ 6\\
\end{tabular}
\end{ruledtabular}
\end{table}

As the simplest strategies $S_{1}$ and $S_{2}$ do not take into account the current level of damage, the failures propagate over the whole network, and the minimum required resources are independent of the response time delay. The $R_{min}$ values are listed in Table \ref{tab:tab1}. Strategy $S_{1}$  demands the highest disposition of resources in scale-free structures. This adverse behavior of scale-free networks arises due to the difficulties in the recovery of hubs and can be eliminated by preferential reinforcement of nodes with high out-degrees (compare the $R_{min}$ values of strategies $S_{1}$ and $S_{2}$).

\begin{figure}
    \begin{center}
    \includegraphics[angle=0,width=0.38\textwidth]{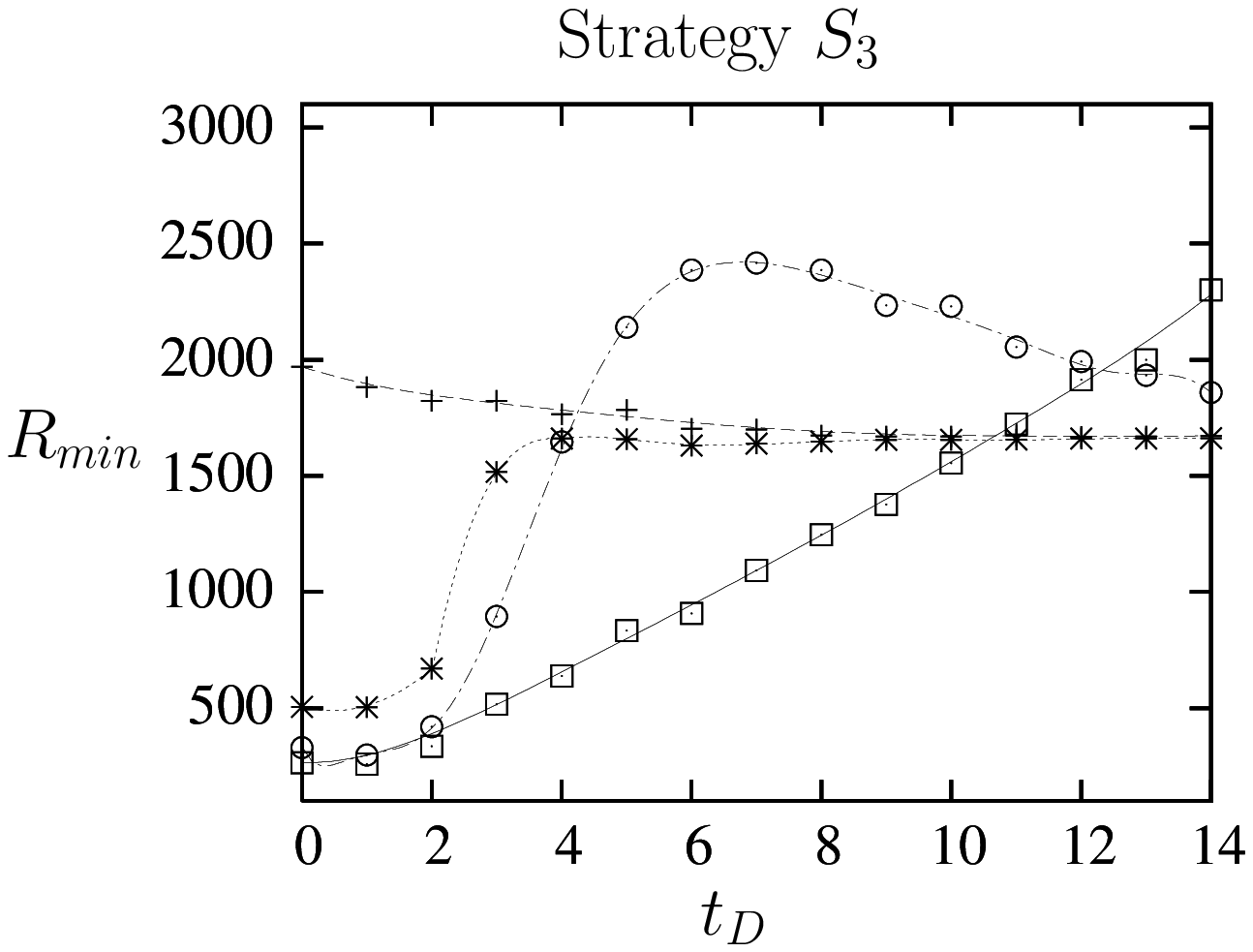}
    \includegraphics[angle=0,width=0.38\textwidth]{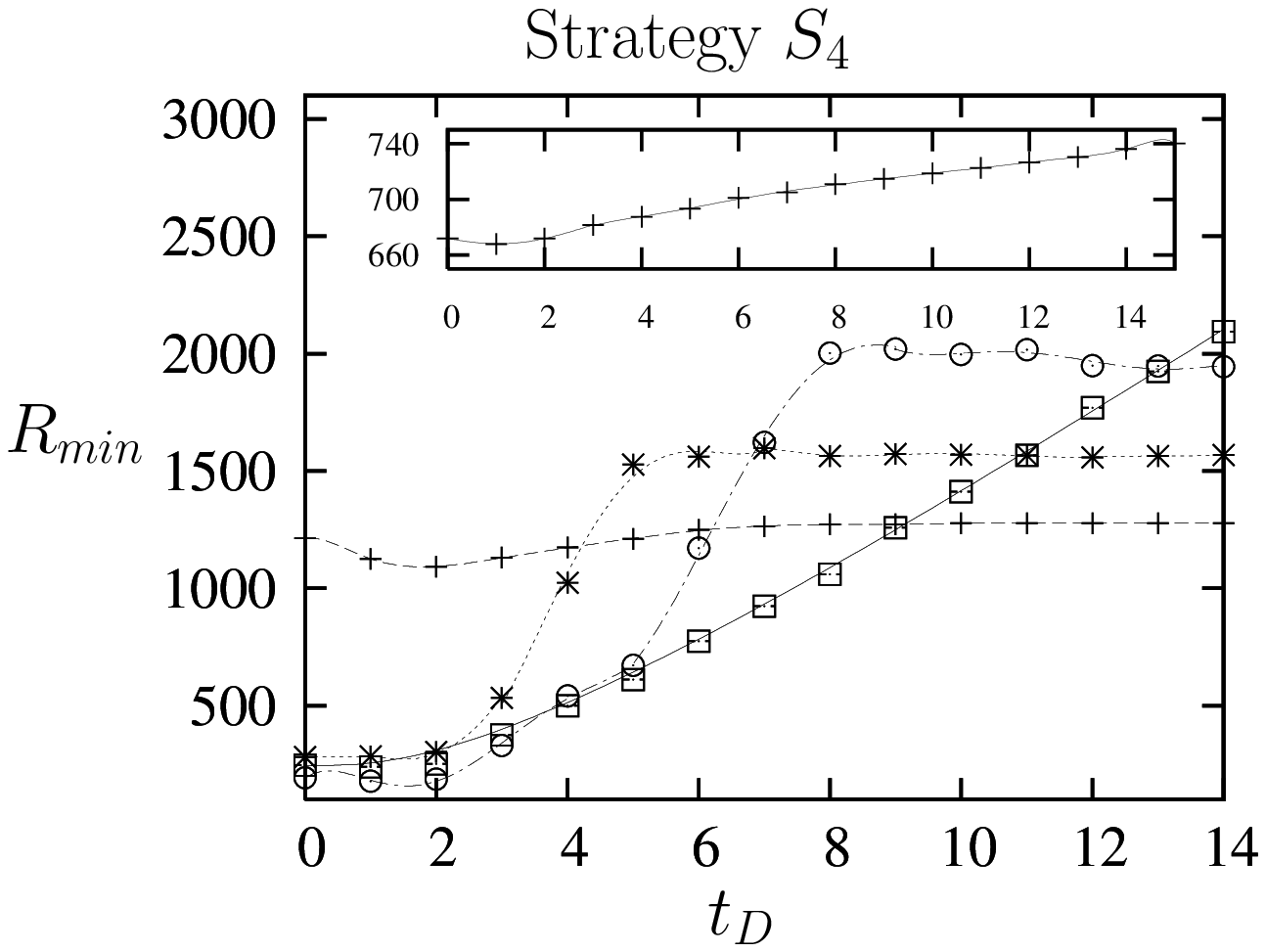}
  \end{center}
  \caption{Minimum quantity of resources $R_{min}$ needed to recover a challenged network as a function of the response time delay $t_{D}$. Squares correspond to bidirectional grid networks, plus signs to scale-free networks, multiplication signs to Erd\"os-R\'enyi networks and circles to small-world networks. The inset shows $R_{min}$ obtained for scale-free networks after the applying of strategy $S_5$ ($k = 0.8$, $q = 0.15$)}.
  \label{fig:Rmin}
\end{figure}

For the damage-based strategies $S_{3}$ and $S_{4}$ (see Fig. \ref{fig:Rmin}) we observe two basic types of behavior: Within the studied range of response time delays $t_{D}$, the  values of $R_{min}$ are either growing, or they stay approximately constant. If they are growing with $t_D$, the resources are sufficient to repair the network before the failures affected the whole network. In the region where $R_{min}$ does not change significantly with increasing $t_D$, damage spreads all over the network. Therefore, the resources required to restore the failure-free state of the network are always the same. 

Our data show the highest spreading velocity for scale-free networks and the slowest spreading for regular grids. The Erd\"os-R\'enyi and small-world networks are somewhere in between and the transition point between the growing and the constant part of $R_{min}(t_D)$ represents the critical value of $t_{D}$ beyond which failures paralyze the complete network.

Small-world networks and, to some degree, scale-free networks as well show a decrease of $R_{min}$ for large values of the response delay time $t_D$, which is surprising (see Fig. \ref{fig:Rmin}, strategy $S_{3}$). This decrease indicates the unbalanced distribution of resources, where there is a surplus of resources in some nodes and a deficit elsewhere. The relationship between the velocity of failure propagation and resources mobilization is crucial for damage-based protection strategies. The spreading velocity is increased by the existence of a small-world effect, which is based on the existence of long-range links (shortcuts). Over these shortcuts, failures spread very fast to distant parts of the network. Consequently, the resources must be distributed over a large area. However, if $t_{D}$ is small, they are deployed less uniformly, because the majority of resources is deployed during the time when only a small part of the network is affected by failures. In such situations, we can find groups of interconnected nodes, which have been less provided with resources. Later on, these nodes require an additional effort to be repaired. In contrast, when $t_{D}$ is large, the resources are distributed more uniformly and the overall demanded quantity of resources is smaller.

In practice, this calls for a precise assessment of the propagation velocity and mobilization rates, which is possible only when the eventually occurring damages can be identified in advance. Taking into account information about the network structure, which determines the possible sequence of failure occurrence, this problem can be significantly reduced (see Fig. \ref{fig:Rmin}, strategy $S_{4}$).

 In order to decrease the spreading velocity in scale-free networks, we suggest to apply strategy $S_{5}$, which stresses the protection of highly connected nodes. Employing a simple heuristic algorithm, we have found values of the parameters $k$ and $q$, which minimize $R_{min}$. The reduction is highest for $k =0.8 $ and $q = 0.15 $ (see Fig. \ref{fig:Rmin}). Although strategy $S_{5}$ utilizes the detailed information about the current damage and network structure, the values of $R_{min}$ for scale-free networks are larger for small values of $t_D$ compared to other networks treated by strategies $S_3$ and $S_4$. On the other hand, for long response time delays $t_D$, the smallest disposition of resources is sufficient to recover scale-free networks.

\subsection{Results for average impact of the different strategies}

\label{subsec:res}

\begin{figure}
    \begin{center}
    \includegraphics[angle=0,width=0.48\textwidth]{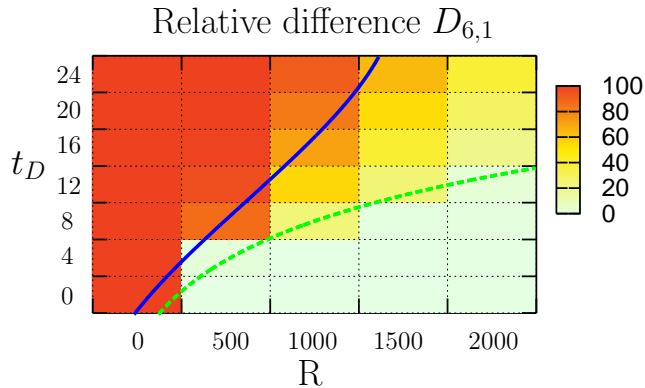}
    \end{center}
    \caption{(Color online) Relative difference in damage $D_{6,1} = \frac{\langle D_{6} \rangle}{\langle D_{1} \rangle} 100 \%$ between the application of the efficient strategy $S_{6}$ and the inefficient strategy $S_{1}$. The dashed line corresponds to parameter combinations for which the  difference between the strategies is 20\%, while the solid line corresponds to a difference of 80\%. The curves have been obtained by simulations using the bisection method.}
\label{fig:data2}
\end{figure}

Before we compare the efficiency of the different disaster response strategies, we will shortly discuss the influence of the strategy parameters on the efficiency of the recovery strategies and take a look at the probabilistic distribution of damage.

A shortage of resources $R$ or a large response time delay $t_D$ can hardly be compensated for, even by sophisticated protection strategies. In the Fig.  \ref{fig:data2}, we compare the typical damage when applying strategy $S_1$ or strategy $S_6$. Strategy $S_6$ was found to be the most efficient one in simulation experiments, while strategy $S_1$ was the most inefficient one (see below). The damage $D_{i}$ related to strategy $S_i$, was quantified by the time integral over the number of destroyed nodes. All results in this subsection are expressed through the average damage $ \langle D_{i} \rangle$, where we varied the initially disturbed node.

Our results show only small differences between the strategies, when $t_{D}$ is large or $R$ is small. However the overall damage of strategies $S_1$ and $S_6$ declines, when R grows and $t_D$ decreases. The superiority of strategy $S_6$ over the strategy $S_1$ is most significant in the region of large resources $R$ and short response delays $t_D$. Thus, improvements in the protection strategy have the highest effect when the response time delay and the disposition of resources for recovery are within reasonable limits, while late response cannot be compensated for even by the best strategies. Similar results have been found for smallpox outbreaks in social networks \cite{Eubank}.

\begin{figure}
    \begin{center}
    \includegraphics[angle=0,width=0.38\textwidth]{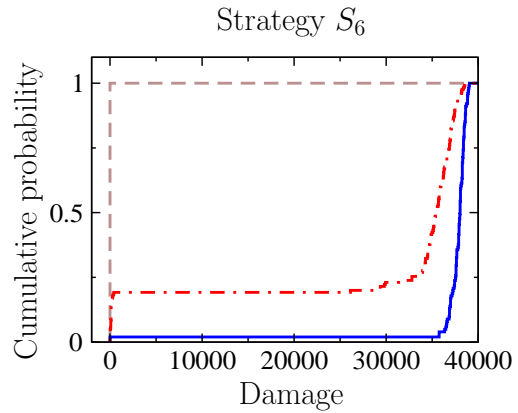}
  \end{center}
  \caption{(Color online) Cumulative probability distribution of the overall damage $\langle D_6 \rangle$ for a sample of numerical experiments for Erd\"os-R\'enyi networks with a fixed disposition of resources $R = 1000$ and different values of $t_{D}$. The dashed line corresponds to $t_D = 0$, the dashed-dotted line to $t_D = 8$, and the solid line to $t_D = 16$.}
  \label{fig:dist}
\end{figure}

A growing response time delay has a strong impact on the distribution of damage. When we fix the amount of resources $R$ and vary $t_{D}$, the damage $ \langle D_i \rangle$ is typically distributed in the way shown in Fig. \ref{fig:dist}. For small values of $t_{D}$, the recovery process is able to repair the network in a very short time (dashed line). For intermediate values of $t_D$ two distinct situations are observed (dashed-dotted): Depending on the initial disturbance and on the random parameters, the spreading is either quickly stopped and the network is recovered. Or, the recovery process is not able to interrupt cascade failure over the entire network, when the number of infected nodes exceeds a certain quantity. For $R > R_{min}$, the system is still repaired, but much later, than for small values of $t_D$. Thus, for intermediate response time delays we can expect a big discrepancy between the damage in the  best and the worst case scenario. This behavior strongly reminds of the initial phase of real disasters, where an apparently irrelevant event like a small social conflict, a thrown cigarette or a delayed disposal of waste can, under similar conditions, either vanish without any significant impact or trigger riots, forest fires or epidemics.

\begin{figure}
    \begin{center}
    \includegraphics[angle=270,width=0.25\textwidth]{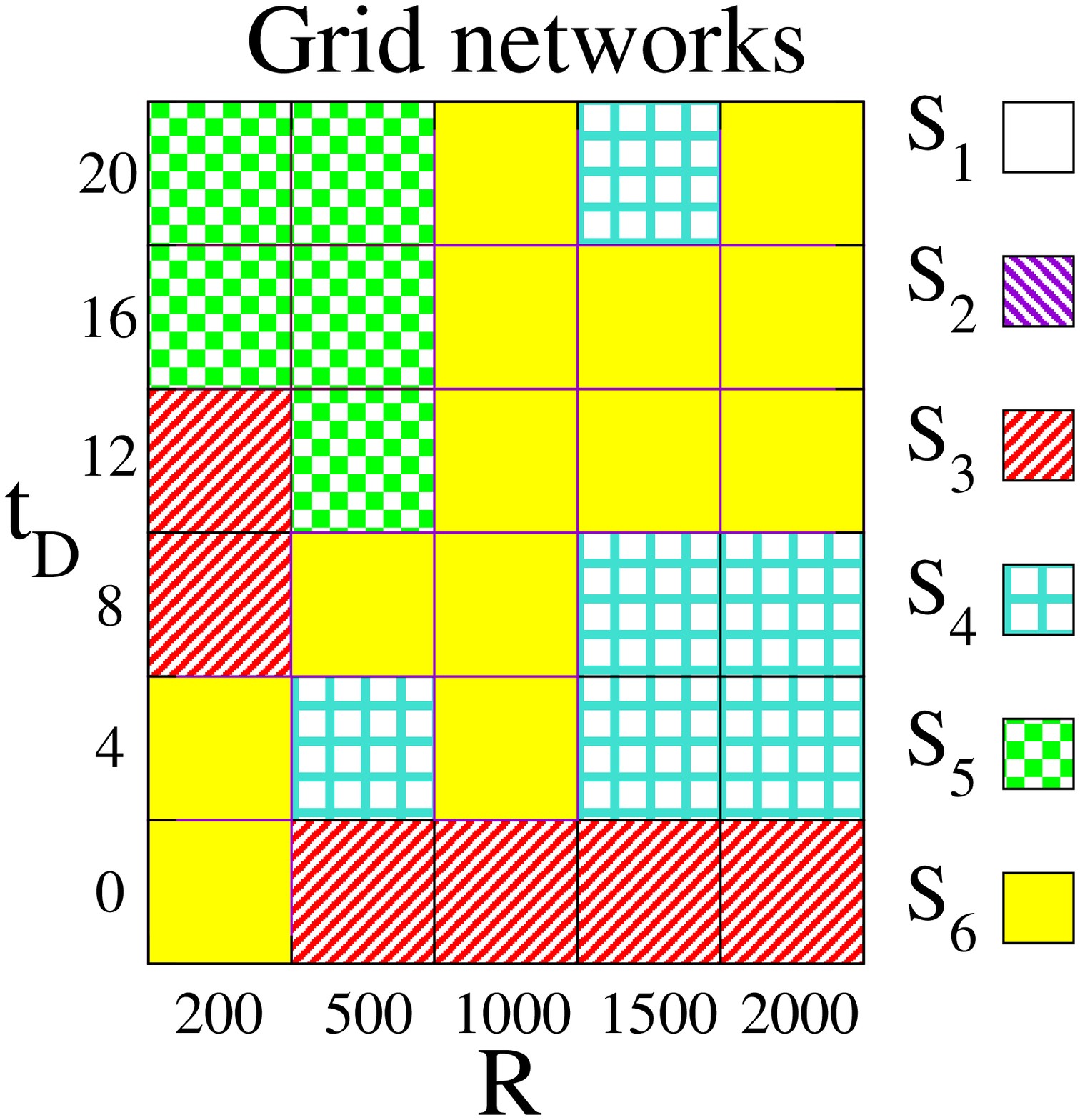}
    \includegraphics[angle=270,width=0.25\textwidth]{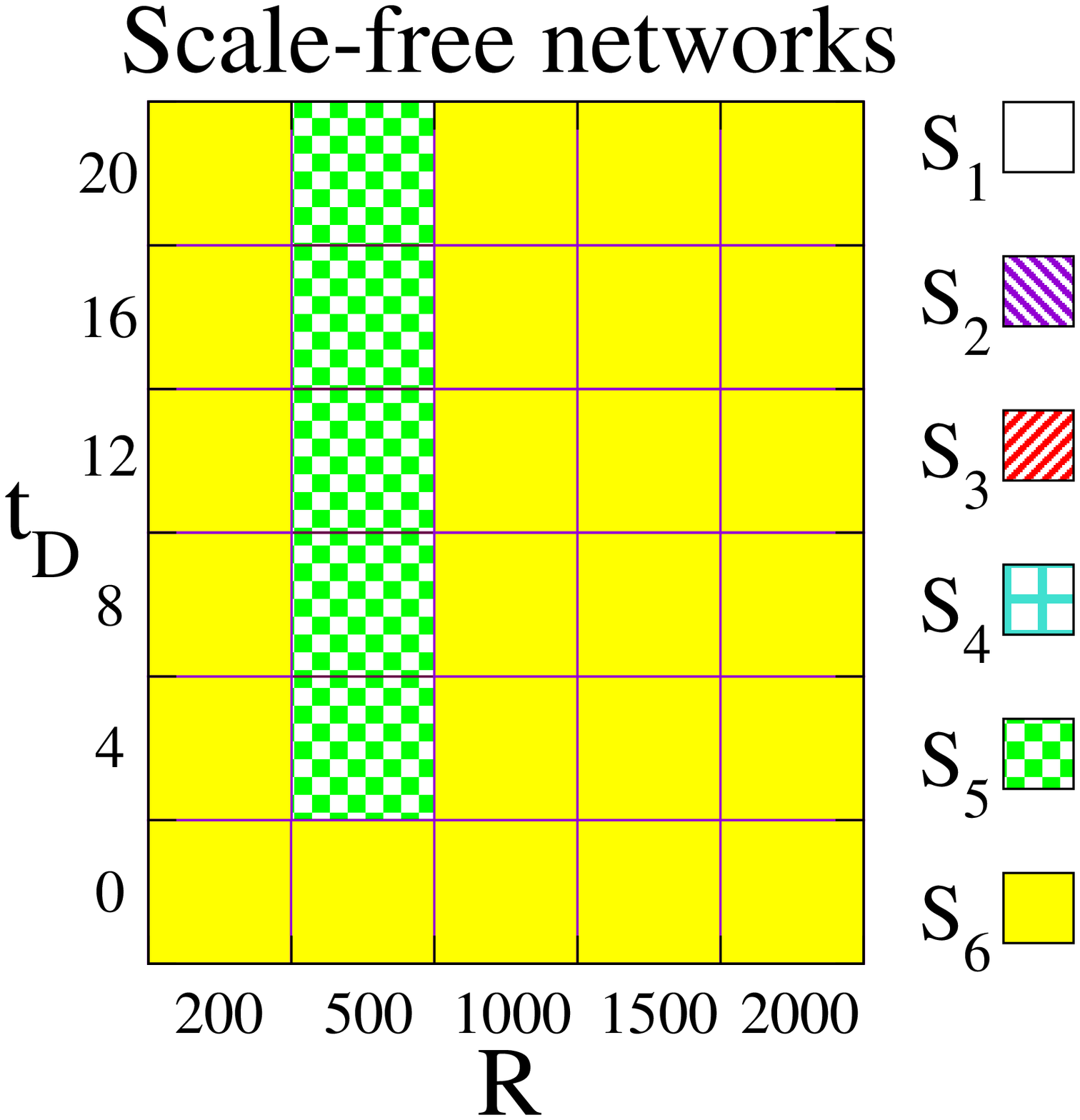}
    \includegraphics[angle=270,width=0.25\textwidth]{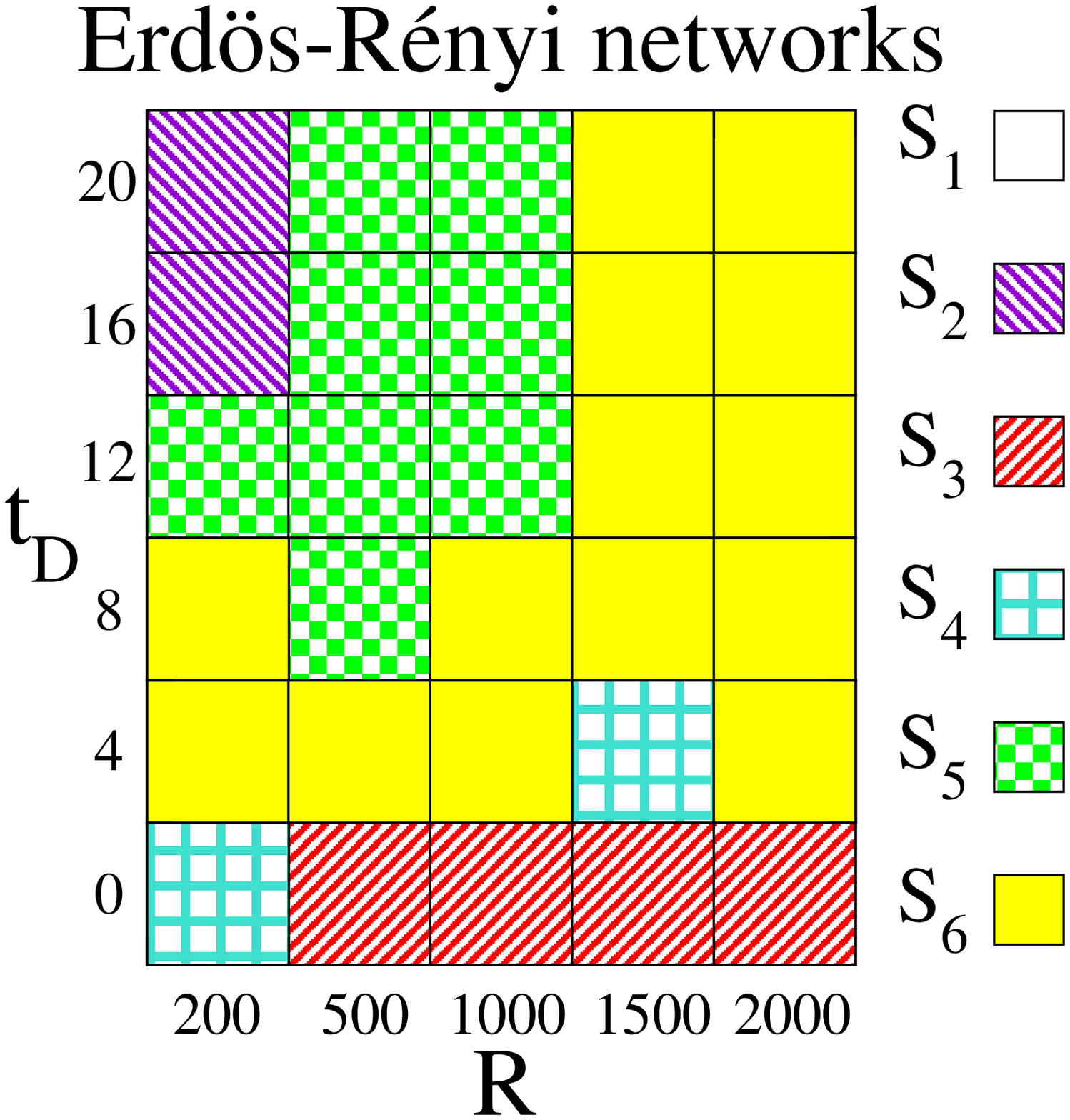}
  \end{center}
  \caption{(Color online) Most efficient strategies of disaster recovery based on the evaluation of average damage.}
\label{fig:comp}
\end{figure}

In order to answer the question which strategies are more proper for which kinds of networks, we have compared the average damage $ \langle D_i \rangle$ for a matrix of parameter combinations (with $t_D \in \{0, 4, 8, 12, 16, 20\}$ and $R \in \{200,500, 1000, 1500, 2000\}$). As the behavior of small-world networks is very similar to Erd\"os-R\'enyi networks, we omitted them in Fig. \ref{fig:comp}. The strategy $S_{5}$ has been particularly suited for scale-free networks to reach the minimum disposition of resources required for network recovery. This strategy is most efficient for values of $R$ close to $R_{min}$. For Erd\"os-R\'enyi, small-world and grid networks, the success of this strategy depends on the respective values of $R$. Strategy $S_5$ is relatively effective, when $R$ is small and $t_D$ is large (note, that for this combination of parameters the differences between the strategies are very small, see Fig. \ref{fig:data2}). However, when $R$ is large, strategy $S_5$ performs poorly, due to the excessive provision of resources to a small group of nodes regardless of the damage. The most universal and also most effective of all investigated strategies is strategy $S_6$. On the other hand, (together with strategy $S_5$), it also requires the most detailed information.

The overall results of our comparison can be summarized as follows: If we have the option to choose whether to orient the disaster recovery strategy at the network structure or at the current damage, then, regular grids with a small spreading velocity are protected best by strategies reacting to the level of damage. In contrast, for scale-free networks it is more effective to take the network structure into account. The choice of the proper strategy for Erd\"os- R\'enyi and small-world structures depends on the response time delay. For short time delays, there is a good chance to reduce  the spreading by preferential protection of damaged nodes, but when the time delay is large and many nodes have already been affected, the damage is minimized by protection of nodes with high out-degrees.

\section{CONCLUSIONS}
\label{sec:con}

Disaster recovery and the operation of inter-connected infrastructures involve an intricate decision making where each action can invoke a variety of hardly predictable reactions. Here the network type plays an important role, and the theory of complex systems and the statistical physics of networks offer powerful methods. These allow one to gain a better understanding of the dynamics of disaster spreading and to derive valuable results how to fight them best.

 In this paper, we have specifically studied the efficiency of several strategies to distribute resources for the recovery of disaster-struck networks. These strategies use information about the network structure  and knowledge about the current damage.  As main parameters, we have considered the overall quantity of resources $R$ and the response time delay $t_D$. By means of simulations, we have determined the minimum disposition of resources, which is necessary to stop disaster spreading  and recover from it. The behavior of scale-free networks was found to be ambiguous. In comparison with other network structures, the highest quantity of resources for recovery is needed in case of small response time delays, while the required disposition of resources is smallest for large time delays.

 When the response time delay and disposition of resources are within reasonable limits, the optimization of protection strategies has the largest effect. Furthermore, strategies oriented at the network structure are efficient for scale-free networks, while strategies based on the damage are more appropriate for regular grid networks. The suitable strategy for Erd\"os-R\'enyi and small-world networks depends on the response time delay. 
In case of short time delays, the damage reduction is higher for damage-based strategies, whereas strategies oriented at information about the network structure are better for large response time delays. Therefore, we expect that the properties of response strategies could be further improved by switching between different strategies in time. This will be a subject of our forthcoming investigations.

\section*{ACKNOWLEDGMENTS}
The authors are grateful for partial financial support by the German research foundation (DFG project He 2789/6-1) and the EU projects IRRIIS and MMCOMNET.


\begin{thebibliography}{10}
\bibitem{footnote}
The function $f(O_{j}) = (aO_j)/(1 + bO_j)$ is motivated by the assumption that the impact of highly connected nodes is distributed among many neighboring 
nodes and, therefore, may decrease with the outdegree $O_j$. A simple linear dependence would be more appropriate in cases, when the spreading depends 
on the transmission of some conserved quantity like, for example, in electrical circuits or in road traffic. However, in some cases the spreading does not 
obey any conservation law, as for the spreading of forest conflagrations or epidemics. Since we do not consider any concrete scenario in this paper,
we decided to use a formula that is more general than a linear function. Other decaying functions in $O_j$ are expected to have qualitatively
similar effects.
\end{thebibliography}
\end{document}